# Wave breaking onset of two-dimensional deep-water wave groups in the presence and absence of wind

Arvin Saket[1], William L. Peirson[1], Michael L. Banner[2], Xavier Barthelemy[1,2]

and Michael J. Allis[1,3]

[1]Water Research Laboratory, School of Civil and Environmental Engineering, UNSW Australia, King St., Manly Vale NSW 2093, Australia

[2]School of Mathematics and Statistics, UNSW Australia, Sydney 2052, Australia

[3]National Institute of Water and Atmospheric Research, Hamilton 3251, New Zealand

**Abstract**

The criterion for the initiation of breaking demonstrated numerically by Barthelemy *et al.* (2015) has been investigated in the laboratory for unidirectional wave groups in deep-water and extended to include conditions of moderate wind forcing. Thermal Image Velocimetry was used to compare measurements of the crest surface water particle velocity ($U_s$) with the wave crest velocity ($C$), as determined by an array of closely-spaced wave gauges. The energy flux ratio $B_x = U_s/C$ that distinguishes maximum recurrence from marginal breaking was found to be 0.840±0.016 in good agreement with the numerically determined value of 0.855. Further, the threshold was found to be robust for different classes of wave groups of distinct characteristic steepness at the breaking threshold. Increasing wind forcing from zero to $U_{\lambda/4}/C_0$=1.42 increased this threshold by 2%. Increasing the spectral bandwidth (decreasing the Benjamin-Feir index from 0.39 to 0.31) systematically reduced the threshold by 1.5%.

**1- Introduction**

Water wave breaking is a dominant forcing process of the upper ocean, inducing strong flow-turbulence-wave interactions and air-sea exchanges (Banner & Peregrine, 1993; Melville, 1996; Perlin & Schultz, 2000; Perlin *et al*., 2013; Peirson *et al.*, 2014). A fundamental and long-standing gap in our understanding of deep-water wave breaking is how to characterise and predict the breaking onset. Many geometric, kinematic and dynamic breaking criteria have been proposed over the past 135 years (Perlin *et al.,* 2013). Geometric and kinematic approaches have failed to determine robust breaking onset thresholds (Perlin *et al.*, 2013).

Song and Banner (2002) proposed a dynamic criterion for two-dimensional waves based on a dimensionless diagnostic group growth rate function $\delta$ as an indicator to predict the onset of wave breaking. They found that the threshold of $\delta_{th}$ in the range of $1.3 \times 10^{-3}$ to $1.5 \times 10^{-3}$ can be used to determine the breaking or recurrence. This was confirmed experimentally by Banner and Peirson (2007) and Tian *et al.* (2008).

This present experimental study is motivated by recent three-dimensional theoretical and numerical studies by Barthelemy *et al.* (2015). Their theoretical work shows that group growth rates, re-cast in terms of energy fluxes, collapse to a much simpler kinematic criterion. Further, Banner *et al.* (2014) have shown that the dominant waves in a group systematically slow down as they pass through the group maximum. Barthelemy *et al.* (2015) show numerically that the onset of breaking occurs once the water surface particle speed at the wave crest ($U_s$) exceeds a set proportion of the speed of the slowing crest ($C$) as it passes through the maximum of a wave group.

Using Thermal Image Velocimetry (TIV) techniques to measure water surface particle speeds at the crests of waves transitioning through a group maximum (the spatial and temporal instance at which group energy density is at a maximum), we critically examine the robustness of this kinematic criterion for group waves in the laboratory of differing group classes, spectral bandwidths and degrees of wind forcing.

**2- Experimental facilities and methods**

**2-1- Laboratory facilities**

The experiments were conducted in the two-dimensional wind-wave tank at the Water Research Laboratory used previously by Banner and Peirson (2007, Figure 1). The flume is 30 m in length, 0.6 m in width and 0.6 m in depth with glass side walls and solid floor. Waves were generated using a computer-controlled, flexible cantilevered wave paddle located on the front side of the flume. The water depth during these present experiments was 0.46 m.

The tank configuration was identical to that used by Banner and Peirson except that a movable wind tunnel of length of 7.5 m was mounted on the tank, with the roof of the tunnel 0.5 m above the still water surface. At the upwind end of the tunnel, an adjustable honeycomb flowguide of 50 mm thickness and composed of 8 mm diameter tubes was installed to establish a uniform air flow within the wind tunnel when air was drawn through the tunnel by a fan at its downstream end. Wind intensity was controlled by varying the fan input voltage. The wind speed was measured on the centreline of the tunnel approximately 4.8 m downwind of the inlet and 0.25 m above the still water level using a pre-calibrated hot probe air velocity meter (Velocicalc model 8347).

A thermal camera was mounted on the wind tunnel roof to observe the tank water surface through a shuttered window 3.6 m downwind of its inlet. The movable wind tunnel could be positioned so that the camera observed the water surface at the location of the group maximum (the locations at which a repetitive wave group has its extreme amplitude). Thermal Image Velocimetry (TIV) was used to measure the horizontal water particle velocities at the crests of waves transitioning through the group maximum. The entire TIV system consisted of an irradiating source, a computer-controlled shutter, the thermal imaging camera and a computer controlling the system components.

Surface irradiation was provided by a pulsed $CO_2$ laser (Firestar TI100) mounted at the centre of the wind tunnel roof and aligned using an adjustable IR range flat mirror. Pulses triggered by the controlling computer were timed to create a sequence of circular heat patches of approximately 4 mm diameter at locations just upstream of the group maximum location.

Surface reflections could potentially damage the thermal imaging camera (Flir T420) used to acquire images of the moving heat patches. Consequently, the computer-controlled shutter remained closed during surface irradiation.

After irradiation, the shutter was opened and the thermal imaging camera was able to observe the water surface vertically from above, capturing 320 by 240 pixel images of the surface at 30 frames per second.

The physical resolution of the thermal imager was approximately 0.66 mm per pixel at the water surface, determined using calibration grids placed within the field of view.

Wave paddle amplitudes corresponding to maximum group recurrence and marginal group breaking were determined by illuminating the tunnel and observing waves through the glass walls. The waves propagating along the tank were two-dimensional. Wave motion was monitored by two linear arrays of capacitance wave probes mounted 50 mm in from each of the tank side walls. To minimise any effect of the probes on the wind flow, the wave probe signal conditioning boxes were mounted outside of the tunnel with 6 mm diameter cables connecting the boxes to the 3 mm diameter, 250 mm long probe frames.

The central probe of each set of wave probes was positioned at the same fetch as the centre of the thermal imaging area. Each probe was fitted with a 200 mm long, 0.2 mm diameter wire element. The wave data was captured using a National Instruments PCI-6225 data acquisition at 1000 Hz sample rate per channel. The probe resolution was 0.1 mm with the linearity of ±0.2 mm over their length.

To measure the crest velocities, 5 wave probes were installed along one tank wall with the spacing of 60 mm. Using the time-series recorded by wave probes, the time of the dominant crest arrival at each wave probe was determined and thereby the crest velocity at the central probe was calculated.

To measure the crest and wave length, seven wave probes were located at the other side of the tank with the spacing of 100 mm. The water level time series captured by the wave probes were interpolated in space to obtain the zero-crossing locations at the time of the wave group maximum and, thereby, the crest length. The local steepness $S_c$ was determined as the ratio of the maximum crest elevation and the crest length at the instant of the wave group maximum.

Fast Fourier Transforms were used to obtain the single-sided wave spectra, the peak frequency ($f_0$), the frequencies associated with the half-peak energy ($f_{min}$ and $f_{max}$) and frequency band width ($\Delta f=(f_{max}-f_{min})/2$). The Benjamin-Feir Index (BFI=$\varepsilon\sqrt{2}/(\Delta f/f_0)$) was used for comparison with other studies (e.g. Janssen, 2003) where $\varepsilon = (k_0^2 <\eta^2>)^{1/2}$, $k_0$ is the characteristic wavenumber associated with the peak frequency and $<\eta^2>$ is the variance of the average surface elevation. Linear wave theory was used to calculate the corresponding wavenumber $k_0$ and linear phase velocity $C_0$ for each wave group.

**2-2- Initial wave group conditions**

The wave packets generated by the paddle were selected to correspond to the cases used by Banner & Peirson (2007). Both a bimodal spectrum (Class 2) and the chirped wave packet (Class 3) were considered in the current study.

The bimodal initial spectrum was defined as:

$$\eta = a_0 \cos(k_0 x) + \varepsilon a_0 \cos\left(\frac{N+1}{N}k_0 x - \frac{\pi}{18}\right) \tag{1}$$

where $\eta$ is the water surface elevation, $a_0$ is the initial amplitude, $N$ is the number of waves in the group and $\varepsilon=0.1$. In the present study values of $N=3$ and $N=5$ were used for the Class 2 wave packets (denoted hereafter as C2N3 and C2N5).

The chirped (Class 3) wave packets were generated using:

$$x_p = a - 0.25 A_p \left(1 + \tanh \frac{4\omega_p t}{N\pi}\right)\left(1 - \tanh \frac{4(\omega_p t - 2N\pi)}{N\pi}\right) \sin\left[\omega_p(t - \omega_p C_{t2} t^2/2)\right] \quad (2)$$

where $x_p$ is the wave paddle displacement, $A_p$ is a reference paddle amplitude, $\omega_p$ is the paddle angular frequency and $C_{t2}$ is the chirp rate of the linear modulation. A paddle frequency $\omega_p$ of 8.18 rad.s$^{-1}$ was used for three Class 3 wave packets with $N$=5, 7 and 9 (denoted hereafter as C3N5, C3N7 and C3N9).

**2-3- Method**

Prior to each experiment, the tank water surface was cleaned by generating waves for approximately 1 hour at the beginning of each day. Any surface slick material was transported to the dissipating beach at the far end from the wave generator. Once the water surface had been cleaned, the wave probes were immersed into the tank for approximately 1 hour to ensure that their signals were stable. The tank surface was inspected regularly to ensure that it was not contaminated with any slick material. Tank water temperatures were recorded before and after each experiment.

The threshold amplitudes for breaking for the five different wave packets described above were determined in the absence of wind. In addition, the thresholds for the C3N7 wave packet were determined with wind forcing applied. In each case, paddle amplitudes were incrementally increased to determine the conditions of maximum recurrence (that is, the maximum paddle amplitude at which no surface rupture was observed anywhere on the water surface) and marginal breaking (the minimum paddle amplitude at which consistent breaking was observed at the point of maximum wave group amplitude) in each case. The fetches of the group maximum were carefully recorded.

Group maximum fetches (and therefore the fetch of marginal breaking) increased systematically with the number of waves in each group. Once the group maximum fetch had been determined, the entire wind tunnel assembly was located over this point and measurement of the local wave characteristics and surface current proceeded.

Measurements were undertaken for the different group classes and wind forcing conditions shown in Table 1. The most challenging measurements were the crest surface velocities at the point of group maxima.

This required ensuring that the position of heat patches coincided with the crest maximum concomitant with group maximum occurrence. Consequently, the sequence of heat patches had to be initially positioned on the surface, allowing for their subsequent wave orbital transport to the immediately vicinity of group maximum. Trial and error was used to achieve this.

---- Figure 1 near here ----

Also, the temperature of the heat patches decreased with time. This required careful minimisation of the measurement duty cycle and careful selection of those thermal image sequences in which the thermal patches remained clearly defined (Figure 1).

By differentiating the heat patch positions with time, a sequence of surface water velocities in the vicinity of the group maximum could be determined. An exemplary sequence for C3N7 in the absence of wind is shown in Figure 2. In this figure, $X$ is the position of each point referenced to the location of crest maxima (the centre of the image) and $t$ is time, referenced to the crest maximum event.

---- Figure 2 near here ----

In Figure 2, the arrows show an ensemble of surface velocity measurements obtained from thermal image records obtained in the vicinity of a sequence of crest maximum events for five wave groups.

To obtain a measurement of velocity localised at the time and location of a crest maximum event, the synchronised thermal imagery and wave probe records were processed as follows. First, thermal patch velocities in immediate spatial proximity of the crest maximum were determined at the time of the crest maximum event. It was found that the duty cycle of the coupled laser/shutter/camera/wave probe system could be synchronised to achieve five velocity measurements surrounding the crest maximum location at the time of the crest maximum.

These five velocities were plotted as a function of distance referenced to the crest maximum position as shown in Figure 3. The results indicated that the maximum surface velocity coincided with the location of crest maxima in each case. A polynomial curve was fitted as shown in Figure 3 to determine maximum water velocity $U_s$.

Due to the framing rate, the absolute time reference of the thermal imagery can only be synchronised with the wave probes with an accuracy of ±17 ms. This uncertainty in synchronisation determines an uncertainty in the crest maximum velocity determined by this process. This uncertainty was evaluated using polynomial fits to the data obtained in the vicinity of crest maximum events and is indicated in Figure 3.

The influence of wind on the onset of wave breaking was investigated for the C3N7 case and followed an identical method. The maximum wind speed investigated was 2.0 ms$^{-1}$. Above this speed, the thermal patches created by the laser could no longer be clearly identified and tracked within five thermal image frames captured through a group maximum occurrence event. Measurements were also undertaken at a wind speed of 1.4 ms$^{-1}$ to verify that any observed trends were consistent. In the absence of forced waves, the TIV technique was used to measure the water surface velocities at a fetch of 3.6 m. These water surface velocities were found to be 0.065 ms$^{-1}$ and 0.078 ms$^{-1}$ at the wind speeds of 1.4 ms$^{-1}$ and 2.0 ms$^{-1}$ respectively.

## 3- Results and Discussion

The wave measurements showed that the wave crests systematically slow down as they approach their crest maximum and subsequently reaccelerate thereafter, as described by Banner *et al.* (2014).

The measured crest speeds, crest steepnesses and crest surface water velocities at the instant of the wave group maximum are summarised in Table 1 for each experimental case. Sample values at both maximum recurrence and marginal breaking define the bounds on the onset threshold and are shown in Figure 4. Table 1 shows averaged values obtained from each data set, with uncertainty expressed as standard error.

------- Figure 4 near here -------

The results show a robust global threshold for the onset of wave breaking of $B_x = U_s/C = 0.840 \pm 0.016$. This compares favourably with the value of 0.855 determined numerically by Barthelemy *et al.* (2015). None of the recurrent groups reach the threshold, while all marginal breaking cases exceed the threshold. This threshold is robust for different types of wave groups and shows no dependency on peak spectral wave numbers. As shown in Figure 4, the characteristic local steepness levels at the threshold of breaking between the Class 2 and Class 3 groups are distinct.

------- Table 1 -------

In each case, it is the crest surface water velocity that plays the dominant role in determining the overall value of this parameter. For all wave groups, across the threshold the crest speed remained almost unchanged between the recurrent to marginal breaking wave condition. In contrast, the surface water velocity increased significantly across the threshold.

The sensitivity of these results has been investigated in relation to two factors: wind forcing and group bandwidth.

As shown in Table 1, the degree of wind forcing has been characterised in terms of the wind speed at an elevation of one quarter of the dominant wave length ($U_{\lambda/4}$) above the mean water surface. For wind forcing $U_{\lambda/4}/C_0$ less than 1.42, the determined breaking threshold remains robust as shown in Figure 5.

------- Figure 5 near here -------

However, more careful examination of the C3N7 data presented in Figure 5 and Table 1 shows that as $U_{\lambda/4}/C_0$ increases from zero to 1.42, there is a slight systematic increase in the threshold in $U_s/C$ of approximately 2.0%. Consequently, wind has a slightly stabilising effect on the underlying wave field.

Group bandwidth has been proposed as an important factor in determining the occurrence of extreme waves and so we have assessed its influence on the threshold. We note that spectral bandwidth can change appreciably with fetch and therefore is not a robust means of characterising group wave fields. In the context of these experiments, spectral bandwidth changed by less than ±5% within a distance of ±3$\lambda_0$ around the group maximum location.

As discussed earlier, the fetch to initial breaking increases as the bandwidth increases. Consequently, there is also a correlation between bandwidth and the envelope growth rate immediately prior to breaking inception (Banner and Peirson, 2007, Figure 5).

The most systematic relationship between breaking onset threshold and spectral bandwidth that emerged from the measurements is shown in Figure 6. As shown, the threshold systematically increases with the BFI. Over the range of group bandwidths considered here, the change in the threshold is only a few percent.

-------- Figure 6 near here --------

**4- Conclusions and Recommendations**

Thermal Image Velocimetry has been used to measure the crest surface water velocity at the crest maximum of freely-propagating, unsteady deep water wave groups in the laboratory. Wave crest speeds were determined using an array of closely-spaced wave gauges at the same instant as the crest water velocity measurements.

A robust energy flux ratio $B_x = U_s/C = 0.840 \pm 0.016$ was found that distinguishes maximum recurrence from marginal breaking. This compares favourably with the numerically-determined value of 0.855 demonstrated by Barthelemy *et al.* (2015). However, the present experimental study encompasses different classes of wave groups exhibiting distinct characteristic steepness at the breaking threshold and was found to be robust.

Increasing wind forcing from zero to $U_{\lambda/4}/C_0 = 1.42$ increased this threshold by 2.0%.

Increasing the spectral bandwidth (decreasing the Benjamin-Feir index from 0.39 to 0.31) systematically reduced of the threshold by 1.5%.

These encouraging results motivate extension of this present work to shallow waters, three-dimensional breaking and field conditions.

**Acknowledgments**

Funding for this investigation was provided by the Australian Research Council under Discovery Project DP120101701. The expert technical assistance provided by Mr. Larry Paice and Mr. Robert Jenkins are gratefully acknowledged.

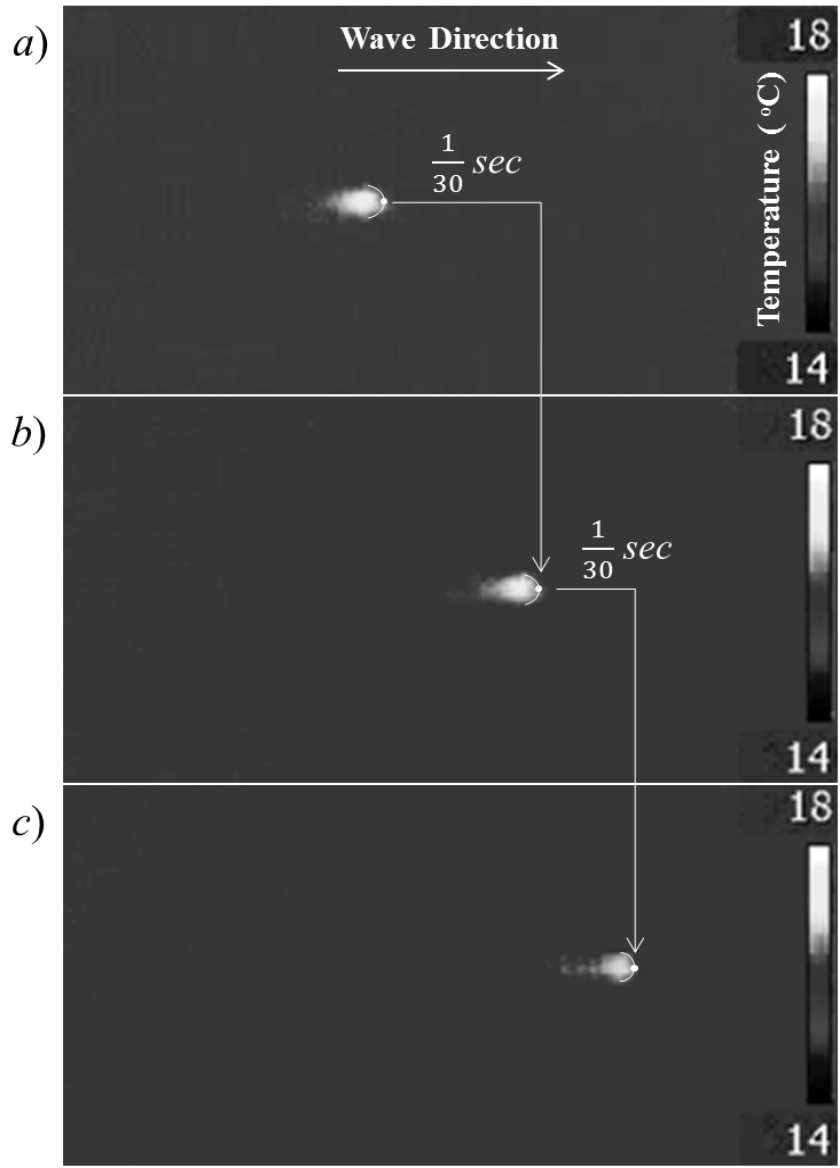

Figure 1. Tracking the hot spot on the wave crest at the location of crest maximum for three successive frames. The camera location is stationary.

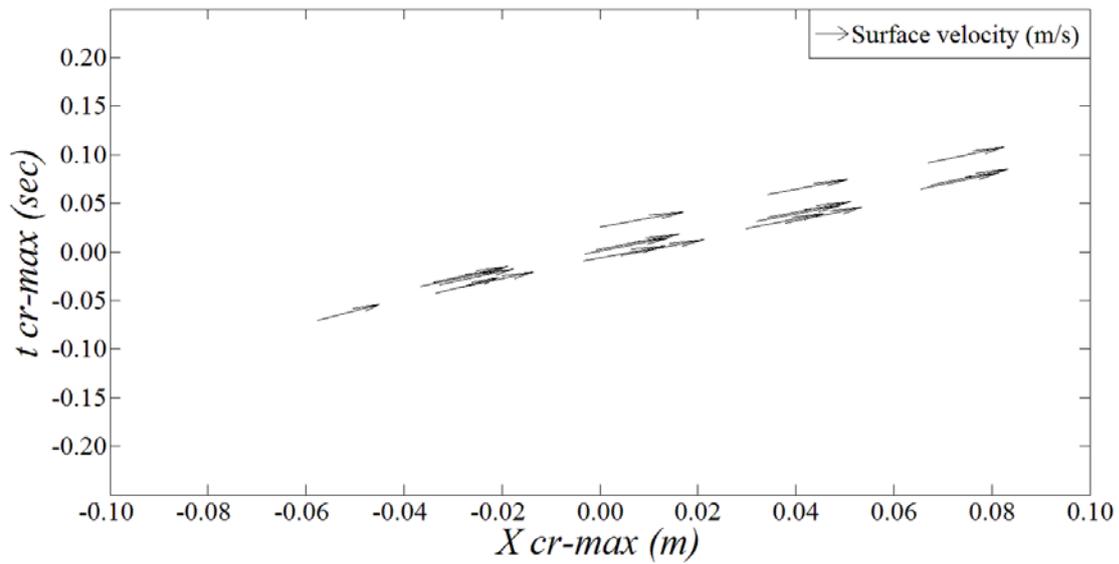

Figure 2. Surface water velocity (m/s) for marginal breaking C3N7 wave group. The arrows represent the surface water velocity, $X$ is the position of each point referenced to the location of crest maximum and $t$ is time referenced to the crest maximum.

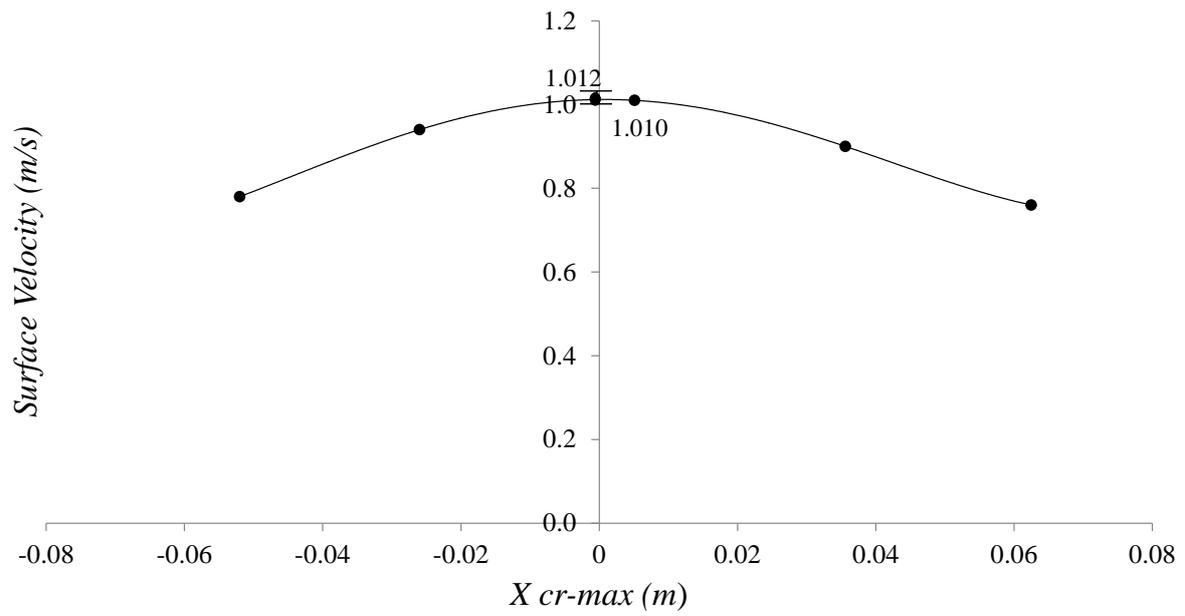

Figure 3. Polynomial curve fitted to surface velocities and the interpolated velocity at the crest maximum for a marginal breaking C3N7 wave group.

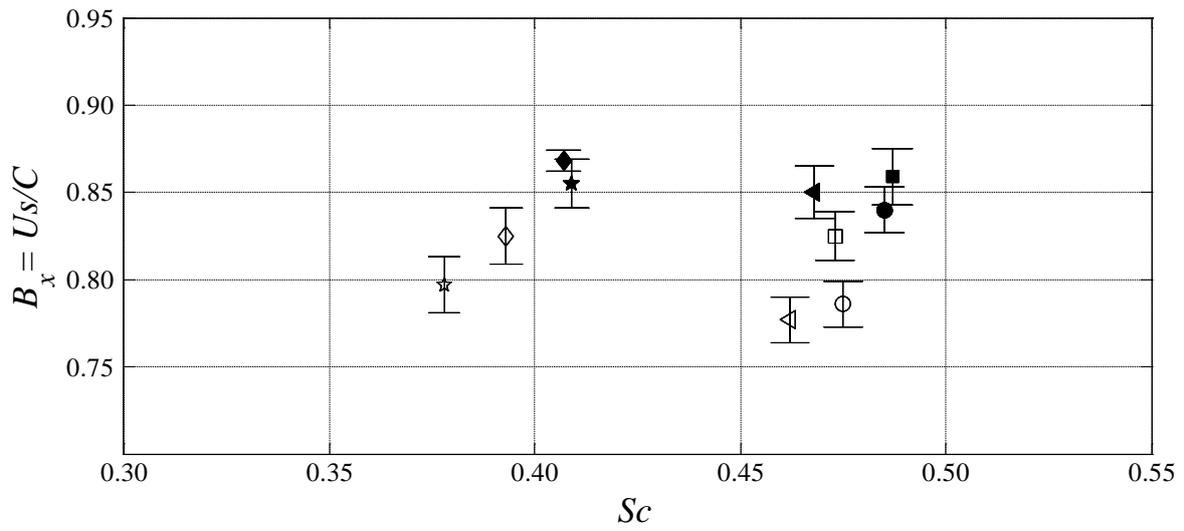

Figure 4. Local wave steepness $S_c$ vs. crest and surface velocities ratio $B_x=U_s/C_{cr}$ for unforced waves, showing C2N3; ◊, C2N5; ☆, C3N5; □, C3N7; ◁, C3N9; ○, with maximum recurrence waves (hollow shapes) and marginal breaking waves (solid shapes).

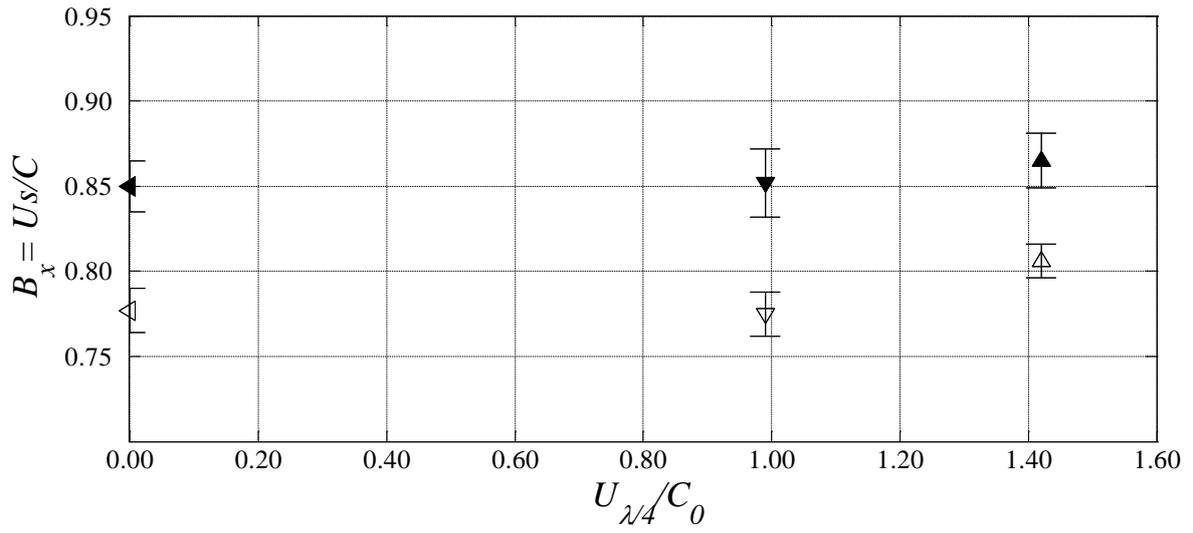

Figure 5. Wind forcing $U_{\lambda/4}/C_0$ vs. Crest and surface velocities ratio $B_x = U_s/C$ for wind forced waves, showing C3N7; ◁, C2N7U1.4; ▽, C3N5U2.0; △, with maximum recurrence waves (hollow shapes) and marginal breaking waves (solid shapes).

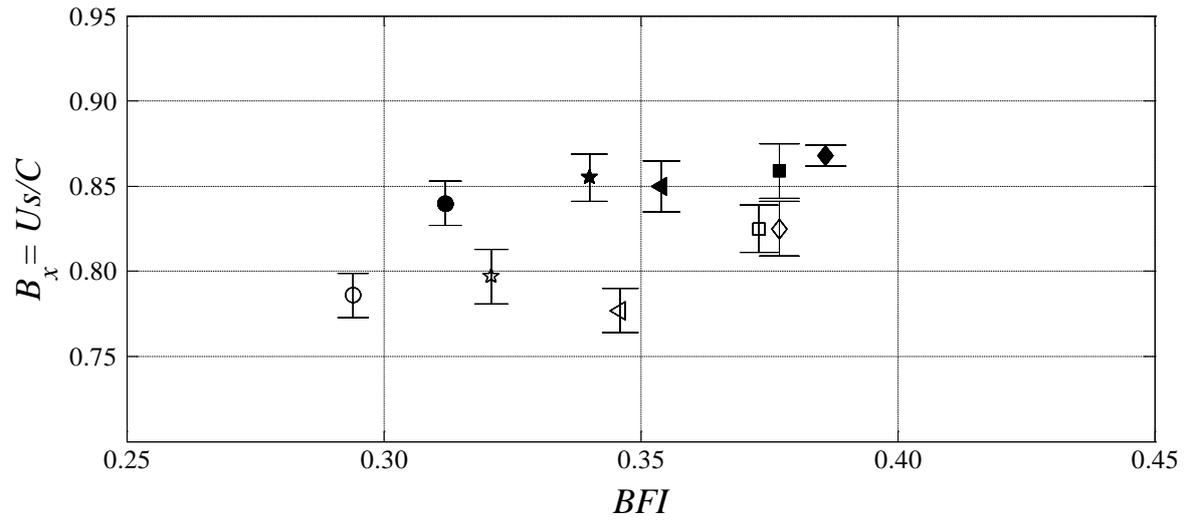

Figure 6. Benjamin-Feir Index *BFI* vs. crest and surface velocities ratio $B_x=U_s/C$ for unforced waves. The shapes are as defined in Figure 4.

Table 1. Wind forcing ($U_{\lambda/4}/C_0$) and the average and standard deviation of measured crest steepness ($S_c$), crest speeds ($C$), crest water surface velocities ($U_s$), energy flux ratio $U_s/C$, peak frequency ($f_0$), frequency band width ($\Delta f$) and Benjamin-Feir Index (BFI) for maximum recurrence and marginal breaking waves.

| Class | $U_{\lambda/4}$ (m/s) | $U_{\lambda/4}/C_0$ | Maximum Recurrence | | | | | | | | Marginal Breaking | | | | | | | |
|---|---|---|---|---|---|---|---|---|---|---|---|---|---|---|---|---|---|---|
| | | | $S_c$ ±0.003 | $C$ (m/s) ±0.014 | $U_s$ (m/s) ±0.017 | $U_s/C$ ±0.014 | $f_0$ (Hz) ±0.001 | $\Delta f$ (Hz) ±0.004 | $\Delta f/f_0$ ±0.003 | BFI ±0.004 | $S_c$ ±0.009 | $C$ (m/s) ±0.014 | $U_s$ (m/s) ±0.019 | $U_s/C$ ±0.016 | $f_0$ (Hz) ±0.002 | $\Delta f$ (Hz) ±0.004 | $\Delta f/f_0$ ±0.003 | BFI ±0.004 |
| C2N3 | - | - | 0.393 | 0.959 | 0.791 | 0.825 | 1.375 | 0.482 | 0.351 | 0.379 | 0.407 | 0.964 | 0.837 | 0.868 | 1.382 | 0.518 | 0.375 | 0.388 |
| C2N5 | - | - | 0.378 | 0.944 | 0.753 | 0.797 | 1.431 | 0.545 | 0.381 | 0.320 | 0.409 | 0.963 | 0.824 | 0.855 | 1.445 | 0.580 | 0.402 | 0.347 |
| C3N5 | - | - | 0.473 | 1.098 | 0.906 | 0.825 | 1.164 | 0.425 | 0.365 | 0.358 | 0.487 | 1.145 | 0.983 | 0.859 | 1.179 | 0.436 | 0.370 | 0.362 |
| C3N7 | - | - | 0.462 | 1.187 | 0.922 | 0.777 | 1.106 | 0.412 | 0.373 | 0.347 | 0.468 | 1.224 | 1.040 | 0.850 | 1.108 | 0.433 | 0.390 | 0.350 |
| C3N9 | - | - | 0.475 | 1.191 | 0.936 | 0.786 | 1.041 | 0.414 | 0.398 | 0.298 | 0.485 | 1.240 | 1.041 | 0.840 | 1.047 | 0.441 | 0.421 | 0.302 |
| C3N7 | 1.40 | 0.99 | 0.459 | 1.237 | 0.959 | 0.775 | - | - | - | - | 0.475 | 1.239 | 1.056 | 0.852 | - | - | - | - |
| C3N7 | 2.00 | 1.42 | 0.449 | 1.255 | 1.011 | 0.806 | - | - | - | - | 0.463 | 1.254 | 1.085 | 0.865 | - | - | - | - |

18